\documentclass[twoside]{article}
\usepackage{fleqn,espcrc2}

\usepackage[dvips]{graphics}

\newcommand{\posi}   {\ensuremath{\mathrm{e^{+}}}}
\newcommand{\elec}   {\ensuremath{\mathrm{e^{-}}}}
\newcommand{\epem}   {\ensuremath{\posi\elec}}

\newcommand{\selec}{\ensuremath{\tilde{\mathrm{e}}}}
\newcommand{\snu}{\ensuremath{\tilde{\nu}}}

\newcommand{\ipb}{\ensuremath{\mathrm{pb^{-1}}}}
\newcommand{\gevcc}{\ensuremath{\mathrm{GeV/c^2}}}

\newcommand{\tanb}{\ensuremath{\mathrm{tan}\beta}}

\title{The Lower Mass Limit on the Lightest Supersymmetric Particle,
using ALEPH data up to 188.6 GeV}

\author{D. Hutchcroft
 \address{Royal Holloway and Bedford New College, University of London,
 London, TW20 OEX, England}
}

\begin{document}

\begin{abstract}
 The Lightest Supersymmetric Particle (LSP) is expected to be stable,
massive and neutral. Direct searches for supersymmetric particles in
the context of the minimal supersymmetric extension to the standard
model have been performed with the ALEPH detector. Using about 175
\ipb\ of data with centre-of-mass energies near 189 GeV a limit on the
mass of the LSP of $M_\mathrm{LSP}>32.3\ \gevcc$ at $95\%$ confidence
can be derived, assuming R-parity is conserved.
\end{abstract}

\maketitle

\section{Introduction}

 The ALEPH detector at LEP collected close to 175 \ipb\ data at a
centre-of-mass energy of 188.6 GeV. Searches in this data for the
decays of Supersymmetric particles have shown no evidence for
supersymmetry. This data can be interpreted as excluded regions in the
MSSM parameter space.  All conventions and notations are consistent
with reference \cite{paper:ALEPH_charg_183}, where the searches for
charginos and neutralinos using data taken at centre-of-mass energies
near 183 GeV are reported. These exclusions extend the previous
results using the ALEPH data given in references
\cite{paper:ALEPH_charg_183,paper:ALEPH_slep_183}.
The results of the searches are interpreted as exclusions in the MSSM
parameter space assuming that R-parity is conserved and the neutralino
is the LSP. Sneutrino masses of less than 42 \gevcc\ are already
excluded by limits on $\Gamma_Z$ \cite{paper:ALEPH_gammaZ_limit}.

\section{Search for gauginos}

 The search for gauginos was performed first under the assumption of
high slepton mass. The visible topologies and energy in gaugino pair
production depend on the decay chain of the gaugino to LSP and on the
mass difference between the gaugino and the LSP. Various topological
searches are used, described in reference
\cite{paper:ALEPH_charg_183}. The selections were reoptimised to give
the best expected limit (in the absence of a signal) for the higher
energy and luminosity.

 In total 25 chargino and 41 neutralino candidates were observed in at
least one of the selections with the expected backgrounds from
standard model processes being 23.0 and 44.3 events respectively. For
the case of high sfermion mass the chargino will predominately decay
via a W$^*$ to the neutralino and the next to lightest neutralino
($\chi'$) will decay via a Z$^*$. For the topologies sensitive to
these cases the number of data events were 10 (4) for chargino
(neutralino) selections with an expectation of 6.8 (5.3) events from
standard model processes. The data sample is consistent with the
standard model expectations and so limits on the production
cross-sections of charginos and neutralinos are derived. Only in the
case of the WW background to the acoplanar lepton search are the
expected backgrounds corrected for in the limits.

\begin{figure}
\resizebox{0.45\textwidth}{!}
{\includegraphics{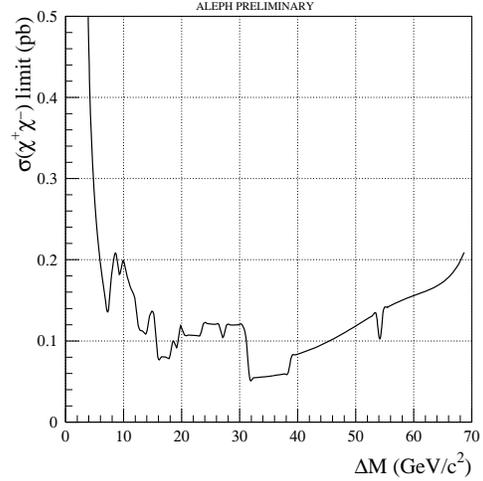}}
\caption[Gaugino cross-section limits]
{The plot shows the $95\%$ confidence level upper limit on the
$\chi^+\chi^-$ production cross-section as a function of $\Delta M =
M_{\chi^\pm} - M_\chi$ for $M_{\chi^\pm}=94\ \gevcc$.}
\label{fig:cs_limits_1}
\end{figure}

\begin{figure}
\resizebox{0.45\textwidth}{!}
{\includegraphics{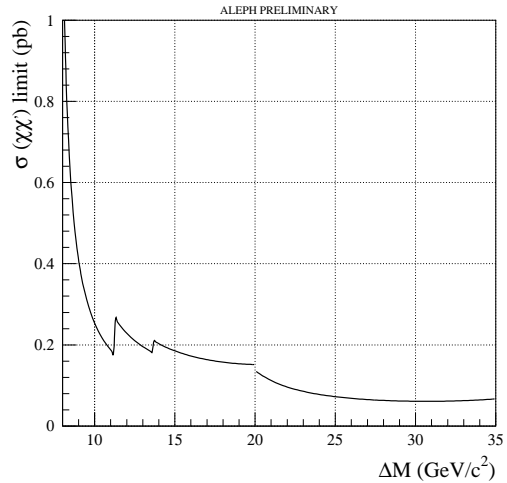}}
\caption[Gaugino cross-section limits]
{The plot shows the $95\%$ confidence level upper limit on $\chi'\chi$
production as a function of $\Delta M = M_{\chi'} - M_\chi$ near the
kinematic limit.}
\label{fig:cs_limits_2}
\end{figure}

\begin{figure}
\resizebox{0.45\textwidth}{!}
{\includegraphics{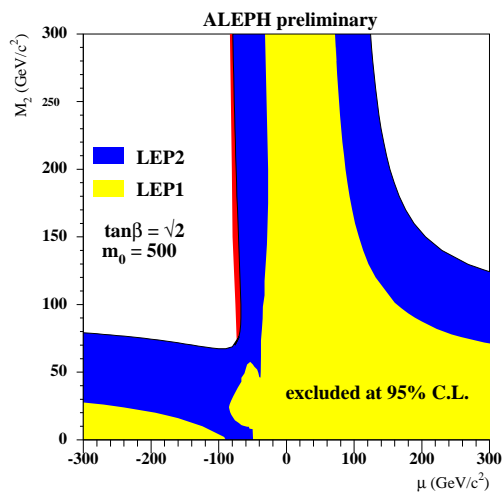}}
\caption[$mu$ v $M_2$ exclusions for $m0=500$ and $m0=75$ ]
{These are the $95\%$ confidence level exclusion contour in the $\mu$
v $M_2$ plane for $\tanb=\sqrt{2}$ and $m_0=500\ \gevcc$. The dark
shading is excluded by chargino searches, the lighter band at negative
$\mu$ is the additional exclusion from neutralino searches.
}
\label{fig:mu_m2_1}
\end{figure}

 The limits for gaugino production close to the kinematic limit are
shown in figures \ref{fig:cs_limits_1},\ref{fig:cs_limits_2} for the
case of heavy sfermions. These limits are shown as excluded contours
in the $\mu$ v $M_2$ parameter space for $\tanb=\sqrt{2}$ in
\ref{fig:mu_m2_1}. Using these exclusions all points in the MSSM
parameter space with neutralino masses of less than 32.3 \gevcc\ are
excluded for any $\tanb$ and $m_0=500\ \gevcc$. 

\section{Search with low slepton mass}

\begin{figure}
\resizebox{0.45\textwidth}{!}
{\includegraphics{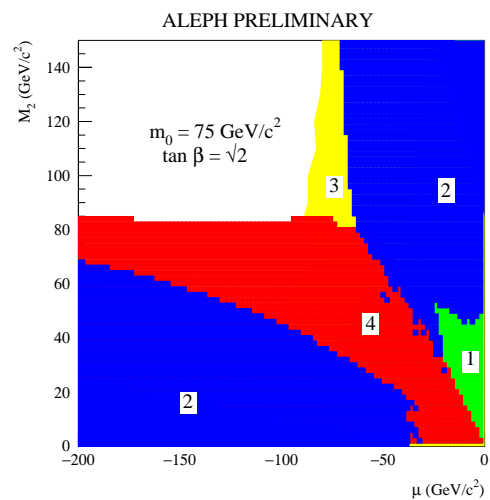}}
\caption[$mu$ v $M_2$ exclusions for $m0=500$ and $m0=75$ ]
{These are the excluded regions in the $\mu$ v $M_2$ plane for
$\tanb=\sqrt{2}$ and $m_0=75\ \gevcc$, the exclusion are from (1) LEP
1 constraints and searches for (2) charginos, (3) neutralinos and (4)
sleptons.}
\label{fig:mu_m2_2}
\end{figure}

The case of low slepton mass is also considered. The chargino and
neutralino pair production cross-sections have a dependence on the
value of $m_0$ due to the s-channel exchange of a $(Z/\gamma)^*$ and
t-channel interference terms for the exchange of a $\snu_e$ or
$\selec$. Also the branching ratio of leptonic decays are enhanced and
invisible modes, for example $\chi'\rightarrow \nu \snu$, become
kinematically allowed. The search for direct slepton production does
however allow additional exclusions and the LEP 1 limit on the
non-standard model contributions to $\Gamma_Z$
\cite{paper:ALEPH_gammaZ_limit} also excludes part of the parameter
space. The combined exclusion for $m_0=75\ \gevcc$ is shown in figure
\ref{fig:mu_m2_2}. Using the combination of these searches all points
in parameter space with $M_\chi<32.3\ \gevcc$ are excluded for all
$m_0$. 

\section{Conclusion}

 The overall limit limit on the LSP mass in the MSSM for the entire
$\mu$, $M_2$, $\tanb$ and $m_0$ parameter space is $M_{LSP}>32.3\
\gevcc$. This limit is set at the point of large $m_0$, $\tanb=1$,
$M_2=54.5\ \gevcc$ and $\mu=-68.3\ \gevcc$, here the lightest Higgs
mass is above 96 \gevcc\ for large $m_A$ and stop mixing. The
sensitivity of the ALEPH Higgs boson searches
\cite{confnote:ALEPH_higgs_189} at 189 GeV does not allow the
exclusion of this point.

\end{document}